# The Case for Dynamic Models of Learners' Ontologies in Physics

Ayush Gupta

Department of Physics

University of Maryland, College Park, MD 20742.

David Hammer

Department of Physics and Department of Curriculum & Instruction,

University of Maryland, College Park, MD 20742.

Edward F. Redish

Department of Physics and Department of Curriculum & Instruction,

University of Maryland, College Park, MD 20742.

Author Note:

Correspondence concerning this article should be addressed to Ayush Gupta, Department of Physics, University of Maryland, College Park, Md 20742. Email: ayush@umd.edu

## Abstract

In a series of well-known papers, Chi and Slotta (Chi, 1992; Chi & Slotta, 1993; Chi, Slotta & de Leeuw, 1994; Slotta, Chi & Joram, 1995; Chi, 2005; Slotta & Chi, 2006) have contended that a reason for students' difficulties in learning physics is that they think about concepts as *things* rather than as *processes*, and that there is a significant barrier between these two ontological categories. We contest this view, arguing that expert and novice reasoning often and productively traverses ontological categories. We cite examples from everyday, classroom, and professional contexts to illustrate this. We agree with Chi and Slotta that instruction should attend to learners' ontologies; but we find these ontologies are better understood as dynamic and context-dependent, rather than as static constraints. To promote one ontological description in physics instruction, as suggested by Slotta and Chi, could undermine novices' access to productive cognitive resources they bring to their studies and inhibit their transition to the dynamic ontological flexibility required of experts.

# The Case for Dynamic Models of Learners' Ontologies in Physics

Development of expertise, including characterizations of novice and expert understanding and models of progress, is a core issue of educational and developmental research. The phenomenon of misconceptions, of novices' persistence in incorrect reasoning despite confuting evidence and instruction, is widely recognized but understood from a variety of perspectives (Brown & Hammer, 2008). Accounts in the literature have varied in their attributions of cognitive structure, from alternative theoretical frameworks (McCloskey, 1983; Carey, 1986) to "knowledge in pieces" (diSessa, 1988; diSessa, 1993a). From an instructional standpoint, these models for understanding expert knowledge and describing novice understanding can lead to widely different recommendations for classroom practice, often at odds with each other. So it is important to examine these different models not just for theoretical interest but because they can have serious implications for how we teach science in the classroom.

In a series of well-known papers, Chi and Slotta (Chi, 1992; Chi & Slotta, 1993; Chi, Slotta & de Leeuw, 1994; Slotta, Chi & Joram, 1995; Chi, 2005; Slotta & Chi, 2006) have proposed an account of a difference between expert and novice understanding as based in *ontological categories*. Their work builds on Keil's (1979), which defines an intuitive ontology as "one's conception of the basic categories of existence, of what sorts of things there are." To think of something as *material*, such as an *object*, for example, is very different from thinking about it as *temporal*, such as an *event*; objects and events are in different ontological categories.

The heart of their perspective is that ontological categories are distinct, stable, and constraining, and that a core reason for novices' difficulties is that they think of concepts in physics, such as heat, light, electric current, and diffusion within the category of *matter* rather than within the category of *processes* (Chi, Slotta & de Leeuw, 1994; Chi, 2005; Slotta & Chi, 2006). On this view, misconceptions result from novices' "commitment to substance-based conceptions" (Reiner, Slotta, Chi & Resnick, 2000), with the implication that instruction of these concepts should steer students toward the ontology of processes and away from the ontology of matter. In recent years, this view has been applied to a variety of traditionally difficult to teach areas of science (Furio & Guisasola, 1998; Chiu, Chou & Liu, 2002; Clark, 2006; Ferrari & Chi, 1998; Johnston & Southerland, 2000; Streveler, et. al., 2006; Venville & Treagust, 1998; Tyson, Venville, Harrison, and Treagust, 1997; Tynjala, 1999; Lee & Law, 2001; Kikas, 2004).

In this article, we challenge the attribution of stable, constraining ontologies and argue instead for the development of dynamic models of both learners' and experts' ontologies. Using examples from everyday speech, student reasoning in classroom settings, and scientific text in peer-reviewed science journals, we argue that (a) novices can and do reason across ontological categories in everyday as well as scientific contexts, and (b) that this is a feature they share with experts, i.e. that expert ontologies are dynamic and context-dependent as well. In particular, for both experts and novices, resources associated with one ontological category are often marshaled productively for understanding concepts in another. For this reason, the instructional strategy of trying to prevent students from applying conceptual resources associated with *matter* could undermine student learning.

The organization of this article is as follows. In Section I, we present an overview of the framework of expert-novice ontologies in physics proposed by Chi and her colleagues. We refer to it as the 'Static Ontologies' (henceforth, SO) model and explain our reasons for the reference. In the following sections, we present evidence of ontological flexibility in everyday language and reasoning (Section II), expert discourse among physicists in published literature and professional settings (Section III), and, finally, in novices' reasoning in physics courses (Section IV).  We show that naive physics is not constrained to the *matter* ontology, and we argue that matter-based reasoning provides essential resources for novices and experts alike to understand concepts in physics.  We close the article with theoretical and instructional implications of the emerging dynamic, context-sensitive nature of expert and novice ontological reasoning in physics.

## THE STATIC ONTOLOGIES PERSPECTIVE

In Chi's and Slotta's account, ontological categories are the "basic categories of realities," of what exists in the world, "such as concrete objects, events and abstractions." (Chi, 2005, p. 163-164).  The categories is hierarchical, starting from the primary categories of *matter*, *processes*, and *mental states* (Chi, Slotta, & de Leeuw, 1994), with subcategories forming three distinct trees.  Under *matter*, for example, there are subcategories of *natural* and *artifacts*; under *natural* there are subcategories of *living* and *non-living*, and so on. The principal indicators of ontological categories are attributes of their members.  "An ontological attribute is one that a category member may plausibly have, but not characteristically nor necessarily has." (Chi, 2005, p. 164).  Attributes are evident in the adjectives and adjectival phrases that a speaker applies to the category member; the researchers refer to these adjectives as *predicates*.

We refer to this account as the "static ontologies" (SO) view because of two central features: (1) entities[1] in the world as well as science concepts correctly belong to a single ontological category at any given level on the trees, and (2) a person's ontological view of an entity is assumed to be committed to a single category, except at moments of radical conceptual change.  The term 'static' thus refers to the context-insensitive (and thus time-independent, except at moments of conceptual change) nature of category membership.

To be clear, by calling the view static, we do no suggest that it rules out the possibility of change; indeed, the framework specifically posits that change is needed for novices to develop expertise.  Rather, we use the word 'static' in contrast to a 'dynamic' account, in which entities in the world may have multiple or ambiguous ontological categorization, and a person's ontological view of an entity is sensitive to context and can vary moment-to-moment.   Moreover, although we argue against some central features of the static framework, we intend the word 'static' to be denotative rather than connotative.  Static models have been extremely productive in physics as well as cognitive science (Ward, 2002, p 68-69.).  The static ontologies model has also been extremely productive in drawing attention to an important aspect of reasoning in physics

## Ontological Categories Are Hierarchical And Mutually Exclusive In The SO Framework

The distinctness of categories at a given level of the hierarchy is one of the key epistemological stances of the model:  An entity cannot belong to both the category *matter* (or any of its subcategories) and to the category *process* (or any of its subcategories).  This also applies to attributes, in that a *matter* ontological attribute cannot apply to a member of *process*[2]:

Categories within a given tree *differ* ontologically from any category on another tree, because they do not share any ontological attributes.  For example, any category of MATTER, such as Living Things or Solids, is ontologically *different* from any category of PROCESSES, such as Naturally Occurring Events.  (Chi, Slotta, & de Leeuw (1994), p. 29.)

And again:

"[A]n hour long" is a predicate that may modify a member of the Event category but cannot be used sensibly to modify any member of the MATTER category, such as a dog.  Thus to say "A dog is an hour long" is anomalous (known as category mistake), because even the negation of that statement ("A dog is not an hour long") is nonsensical.  On the other hand, if a member of a category is predicated by an attribute from the same ontology, then, at worst, the statement is simply false (e.g., "A dog is purple"). [emphasis as in original] (Chi & Slotta, 1993, p. 252)

Chi and Slotta similarly expect concepts in science belong to one and only one category at any level of the hierarchy.  For example, heat, light, electric current, and evolution are all *processes*, and it would be a category mistake to apply any matter-like attribute when describing them.  Hierarchical organization and mutual exclusivity applies also to an individual's ontological knowledge: "People's conceptions of entities in the world should correspond to such ontological distinctions" (Chi, Slotta, & de Leeuw, 1994, p. 30).

The psychological reality of ontological distinctness is crucial to the story of conceptual change in this framework:

In sum, ontological distinction underlies our notion of conceptual change: Conceptual change occurs when a concept has to be re-assigned to an ontologically distinct category (*across trees*).  This definition specifies clearly what a distinct ontological category is, by means of the non-overlapping ontological attributes.  (Chi, Slotta, & de Leeuw, 1994, p. 31)

## The SO Framework Assigns Particular Process Subcategories As The Single Normative Category For Many Science Concepts

Chi and Slotta first described many difficult to learn science concepts, including heat, light, electric current, and evolution, as belonging to a specific subcategory of *acausal interactions* (Chi & Slotta, 1993), later refining and renaming that category as *constraint-based interactions* (Chi, Slotta, & de Leeuw, 1994), and again, *emergent processes* (Chi, 2005):

[A]n electric current is neither MATTER nor properties of MATTER, but a PROCESS that is fundamentally constraint-based and has no causal agents.  The same applies to the concept of heat, force, and light, which are all entities whose veridical (metaphysical) conception belongs in the Constraint-based Interaction category.  (Chi, Slotta, & de Leeuw, 1994, p. 31)

In contrast, science concepts that are easier to learn are in the *direct processes* subcategory (Chi, 2005).  *Direct* and *emergent process* categories are "fundamentally distinct" on the basis of differences in the causal mechanisms.  For example, Chi (2005) explained diffusion (of a blue dye in water) and blood circulation as examples of emergent and direct processes respectively:

> "Basically, these two processes differ in the mechanism that causes the flow - that is how the structure (or behavior or function) of the components causes the pattern to occur.  In one case, blood circulation, the nature of the aggregate components or their constituents (such [as] the structure of the chambers of the heart and the strategic placement of the valves in the heart) are directly causing the global pattern of flow, such as its direction and speed.  Therefore, this type of processes can be referred to as direct . . .
> In contrast, for a process such as diffusion, neither the aggregate components themselves (the dye and the clear liquids) nor their constituents (dye and water molecules) are directly causing the global flow pattern to occur.  Instead, the mechanism of the flow must be explained in terms of the collective interactive outcomes of all the constituent components (both dye and water molecules) such that neither an individual component (such as the dye liquid) nor a group of individual constituent components (such as all the dye molecules) cause(s) the global pattern (or the dye to flow). Therefore, the mechanism of flow in a diffusion kind of process is nondirect and can be referred to as emergent. (Chi, 2005, p. 174).

In other words, within the SO framework, anyone who understands diffusion correctly should treat it only as an *emergent-process* (with emergent process attributes) while blood circulation should be treated as a direct process.  Similarly, the veridical nature of heat, light, and electric current are also *emergent process*-like and it would be an error to apply any *direct-process* or *matter* attribute to them.

## The SO Framework Attributes Student Difficulties To Robust Mis-Categorizations Of Concepts

By the SO account of ontological knowledge, persistent student difficulties arise from students' mis-categorizing concepts.  That is, students place concepts and reason about them from the wrong ontological category:

> Many robust misconceptions can be interpreted as a mismatch between conception and reality at an ontological level, rather than (and in addition to) at the concept-specific and theory-specific level.  In this view, robust misconceptions are mis-categorizations across ontological boundaries in that a member of one ontological category is misrepresented as a member of another ontological category." (Chi, 2005, p. 164).

Specifically, SO theorists claim that students tend to categorize science concepts as matter, and so give them attributes of matter:

> For example, students tend to think of heat either as a kind of substance that objects may contain (e.g. hotness) or as concrete objects (e.g. heated molecules, both hotness and heated molecules are types of entities) rather than a process (of movement of atoms or molecules within an object)... We have empirical evidence to support this type of miscategorization of processes-as-substance for concepts such as electric current, heat transfer, and light (Slotta, Chi & Joram 1995)." (Chi, 2005, pp. 187-188).

In addition, when students do think of the concepts as processes they tend to think of them as direct rather than emergent: "Emergent processes are particularly troublesome for students to understand because they misattribute the concept's ontological nature, either as a material substance or as a kind of direct causal process" (Slotta & Chi, 2006, pp. 263-264).

In this view, students' misconceptions are persistent because ontological commitments are difficult to change. Students' lack of understanding of emergent processes, and their "innate predisposition to interpret all processes as a direct kind" (Chi, 2005, p. 186), prevent the appropriate ontological shifts:

> Misconceptions result from commitments to an inappropriate ontology. … Unfortunately, once an ontological commitment is made with respect to a concept, it is difficult through any stages of mental transformation to change one's fundamental conception from a substance to a process…Thus ontologically misattributed concepts would require an extraordinary process of conceptual change. (Slotta & Chi, 2006, p. 263).

As we discuss further below, Chi and Hausmann (2003) acknowledge instances in which novices can and do change their ontological commitments. For example, they note, novices can easily reclassify whales from the ontological category of "fish" to that of "mammal," moving from one branch to another on the ontological tree of *living things*. They maintain, however, that student difficulties in learning science largely reflect "situations in which an ontological shift is necessary, but people find it extremely difficult to do. This is the case when students misunderstand concepts such as heat transfer, electricity, natural selection" (Chi & Hausmann, 2003, p. 8).

## The SO Framework Interprets Verbal Attributes (Predicates) As Ontological Markers

The empirical evidence of categorization is in verbal predicates used by research subjects (Slotta, Chi & Joram, 1995; Slotta & Chi, 2006; Reiner, et. al., 2000). That is, as in Keil's work (1979), the predicates students use to describe concepts provide evidence of their ontological categorization:

> For example, if a participant said, "The current *comes down* the wire and *gets used up* by the first bulb, *so very little* of it *makes its way* to the second bulb," then these four (italicized) predicates were taken as evidence that participants conceptualized current as a substance-like entity with attributes of (a) "moving," (b) "can be consumed," (c) "can be quantified," and (d) "moves," respectively. [emphasis as in original] (Slotta & Chi, 2006, p. 265).

The following is another of their examples, on parallel bulbs in a circuit:

C2: Um, we can say that the bulbs will all come on at the same time, because you're completing the circuit of electricity which runs through each bulb. Like, the battery puts out a certain amount of electricity when the switch is connected, and it gets spread out through each bulb, but if you added up all those electricity within each bulb or wire, it would add up to the whole amount of electricity put into it. (Slotta & Chi, 2006, p. 282)

Slotta and Chi (2006) considered the use of predicates *runs through*, *puts out, a certain amount*, *gets spread out*, and *added up* as evidence that the student categorized current as *matter*.
In what follows, we apply this method of predicate analysis to data from everyday, classroom, and expert discourse to challenge the SO result.

## The SO Framework Implies Instruction Should Try To Prevent Matter-Based Reasoning About Scientific Concepts

Within the SO view, the naïve physics ontological view cannot be refined to the expert view without a restructuring. The implication is that instructors should avoid using or allowing students to use analogies or language associated with the "wrong" category:

This view of learning physics suggests that it is not possible to refine or develop intuitive knowledge to the point that it becomes the veridical physics knowledge; entities on separate ontological trees cannot be merged, because they cannot inherit each other's attributes. For example, the danger in using the analogy of flowing water to instruct about electrical current is that students will continue to assimilate newly taught information about electrical current into the ontological class of MATTER. If students assimilated new information about electric current into the Liquid subcategory, the concept might also inherit properties such as "has volume," "occupies space," and other ontological attributes of the MATTER category. [emphasis as in original] (Chi & Slotta, 1993, p. 256).

Or, more recently:

[T]eachers should not try to "bridge the gap" between students' misconceptions and the target instructional material, as there is no tenable pathway between distinct ontological conceptions. For example, students who understand "force" as a property of an object cannot come gradually to shift this conception until it is thought of as a process of interaction between two objects. Indeed, students' learning may actually be hindered if they are required to relate scientifically normative instruction to their existing conceptualizations. (Slotta & Chi, 2006, p. 286)

The researchers have emphasized repeatedly the need to refrain from any description that captures the 'incorrect' ontology, such as using blood flow as comparison to flow of current in a circuit:

[O]ur research suggests that instruction should stress the basic ontological characteristics of the concepts, targeting students' existing conceptions indirectly by carefully avoiding

any language, analogies, or phenomenon that might otherwise reinforce the substance-based view. (Slotta & Chi, 2006, pp. 286-287)

## The SO View Has Influenced Research On Conceptual Change

The SO framework has had a substantial impact on physics, chemistry, and biology education research.  In this subsection, we discuss several examples of how others have interpreted and made use of the SO framework.

Tyson, Venville, Harrison, and Treagust (1996) proposed a multi-dimensional model of conceptual change, in which development of expertise involves progress from epistemological, social/affective and ontological perspectives.  The ontological  dimension of their model is derived from the SO view.  Venville and Treagust (1996) then applied this view in developing an instructional intervention in an 11[th] grade biology class "to change the students' view of cell membranes from a 'matter' category to a 'process' category" (p. 312).  Subsequently, Venville and Treagust (1998) applied it to explore students' conceptual change in genetics.  In chemistry education, Chiu, Chou, and Liu (2002) adopted "Chi's ontological categories of scientific concepts to investigate how students perceive the nature of chemical equilibrium" (p. 689).  The authors viewed chemical equilibrium as correctly belonging to the category of 'constraint-based interactions'[3] while novices in their study often used matter-based predicates to explain and describe it.  Lee and Law (2001) applied the SO framework to studying and developing a teaching strategy to promote students' conceptual change about electricity.  Kikas (2004) drew on this framework to account for misconceptions in physics.  And in engineering education, Streveler et. al (2004) used the framework to show that even advanced engineering undergraduates may have matter-like ontological view of concepts such as force and voltage.

In each of these examples, researchers have drawn from and contributed to the SO framework.  All have attributed students' misconceptions to their commitment to the *matter* ontology; all have drawn the implication that instruction needs to disrupt or suspend these incorrect commitments and re-orient students to the correct ontologies:

What is important here is that if students do think of concepts as being in different ontological categories than those in which scientists put them, then, in those cases, conceptual change to scientific conceptions much involve ontological changes in the students' cognitive structure (Chi et al, 1994).  From this point of view, students have to stop thinking of concepts like heat, light, force, and current as material substances. (Tyson, Venville, Harrison, & Treagust, 1996, p. 399-400).

And:

Chi et al. (1994) suggested that the most difficult changes are those that involve ontological re-categorization.  Their theory has three important claims. First... all entities in the world belong to three primary ontological categories.... Second, many scientific concepts are 'constraint based interactions' (CBI), which is a subcategory of processes in which a defined system behaves according to the principled interaction of two or more constraints.  Third... learning in [many] instances requires a shift in the concepts' ontological status from the matter category to the process category.  Chi et al. (1994) further suggested that the major barrier to conceptual change relates to the difficulty in making the shift between two distinct ontological categories. (Lee & Law, 2001, p. 112).

Before we continue, it is important to note that our critique focuses on the SO perspective. That perspective has primarily been advanced by Chi, Slotta, and their colleagues, but it is important to emphasize that the framework has a life of its own in the research community.  In fact, as we noted above and revisit below, Chi and Hausmann (2003) differs in some respects from the SO view represented elsewhere.  Our purpose, however, is to challenge the SO perspective as it appears in the literature, which we hope to have documented sufficiently.

To be clear, too, we believe the SO work has been useful, in particular in directing researchers' attention to patterns of reasoning associated with different ontological categorizations.  However, we disagree with the account of ontologies as static that their work began and others have taken up.  Most important, we argue below, instruction that seeks to prevent students from reasoning in ways based on their experiences of matter or of "direct processes," rather than to help them refine and build from that reasoning, is likely to be detrimental to their learning physics.  Much of our motivation for writing this article is that the SO view and its implications are being promulgated in the literature and in some places are having an influence on science teaching.

In what follows, we present arguments and evidence that challenge the SO characterization of expert and novice ontologies in physics.  In Section II, we present evidence from everyday speech to argue against the distinctness of ontological categories in general.  In Section III, we analyze expert discourse in scientific literature as evidence against the idea of a single, veridical category for science concepts.  Finally, in Section IV we analyze novice reasoning to argue against the SO characterization of novices as committed to the *matter* ontology and lacking knowledge of *process* category or the *emergent-process* subcategory.

## ANALYSIS OF EVERYDAY LANGUAGE REFUTES ONTOLOGICAL DISTINCTNESS

Modern psycholinguistic understanding of categorization in everyday language does not support the idea of ontological distinctness as a cognitive property that is the central feature of the SO perspective.[4]

There has been extensive study of categorization in psychology and linguistics over the past few decades. (See Sternberg & Ben-Zeev, 2001 chap. 3, and Evans & Green, 2006, chap. 8 for good surveys and many references.)  Many of the results are complex, and much controversy still exists, but there are some well established results that inform research on ontological categorization.  These arise from the main research thrusts of (1) embodied cognition and (2) cognitive blending.

The idea of embodied cognition is that our cognition is based and built directly on our physical perceptual experience, such as when someone says "I'm feeling down today" or "Lighten up!" (Lakoff, 1987; Lakoff & Johnson, 2003; Evans & Green, 2006)  Cognitive blending provides descriptions of how distinct mental ideas can be combined to yield more complex mental representations (Fauconnier & Turner, 2002, and articles in *The Journal of Pragmatics*, October 2005).  Together, they provide a road map for constructivism – how we build richer and more complex ideas out of straightforward physical experience and perception.

From these accounts, we come to understand concepts that fall in the SO category of *mental states* through metaphor to physical experience.

Two further results of this general framework and of research into category formation[5] are that (1) perceptions of categories can be strongly context-dependent; (A chicken is not a prototypical bird, unless you are in a cooking school – where having a tray of otherwise canonical birds such as robins brought in would be unexpected, even shocking) and (2) people easily generate categories on the spot to deal with a novel situation (such as "objects one would take out of one's burning home") (Barsalou, 1983; Barsalou, 2005).

In this section, we give evidence that such cross-categorical reasoning is commonplace in everyday reasoning and speech.

## Everyday Thinking And Language Often Straddle Ontological Categories

Research in linguistics over the past three decades demonstrates that everyday language is rife with metaphors that cross ontological categories. Lakoff and Johnson (2003) cite examples of familiar expressions and make explicit the underlying metaphors, such as:

"He *cracked* under pressure" (MIND AS BRITTLE OBJECT)
"We are trying to *grind out* the solution to these equations" (MIND AS MACHINE)
"We need to *combat* inflation" (INFLATION AS ENTITY)

It is not difficult to find other examples:
I have a cold - I just can't *shake* it (COLD AS ENTITY)
Man, that party was *massive*! (QUANTIFYING PARTY AS MASSIVE OBJECT)
At the end, everyone *meets* death. (PERSONIFICATION OF DEATH)
Don't keep your anger *bottled up*. (EMOTION AS SUBSTANCE)

One reason ontological metaphors are easily and naturally understood because the metaphorical models that they embody are culturally ingrained (Lakoff & Johnson, 2003). We add that everyday reasoning is sufficiently flexible to handle new metaphors. Thus an ontologically novel expression such as, "[J]ust because we have the technology – and a 24/7 font of information – does not mean we have to infuse it into every last molecule of available time" (Wood, 2007, p. 26), is not considered nonsensical, although it employs a matter-like description of time (and of technology). Adults have little difficulty understanding the meaning of Secretary of State Condoleeza Rice's statement about the Middle East, "[T]he region is just littered with broken ceasefires," (Rice, 2006) despite, again, the use of matter-like predicates applied, in this case, to ceasefires. In other words, in everyday language it does not seem nonsensical for us to apply matter predicates (*infuse*, *litter*, *broken,* and *quantification*-in-molecular-units) to non-material entities (technology, ceasefire, time). In most uses of common metaphors, people do not even notice the ways everyday thinking crosses ontological categories, let alone resist it.

In fact, Lakoff's & Johnson's contention is stronger. They claim that "our experiences with physical objects (especially our own bodies) provide a basis for an extraordinary wide variety of ontological metaphors, that is, ways of viewing events, activities, emotions, ideas, etc., as entities and substances," (Lakoff & Johnson, 2003, p. 25). They support this claim with numerous examples, such as those above, of metaphors based in a matter ontology that support reasoning about processes and mental states.

In other words, Lakoff's and Johnson's work shows crossing ontological boundaries is *essential* for ordinary cognitive development.  The apparent "mis-categorization" of events and emotions as objects helps rather than hinders our understanding of those events and emotions.

Emotions, processes, and objects can share characteristics and verbal predicates.  For example, some events and processes are bound in space like objects; mental states can be bound in time like events, and some objects (living things) can be bound in time as well.  While it may be nonsensical to say that a dog is an hour long, the sentence, 'Our dog lasted 14 years' is perfectly understandable.  There may be ontological categories with mutually exclusive properties, but matter, processes, and mental states evidently can share attributes and predicates without problems.

One response to this could be that while prevalent in everyday speech, making cross-categorical connections is cognitively less viable or less appropriate in the context of scientific reasoning (Chi and Hausmann, 2003).  In the following sections, we present arguments and evidence that not only is such ontological flexibility appropriate in scientific contexts, but novices and experts alike can straddle ontological categories when reasoning about science concepts.

## EXPERT UNDERSTANDING STRADDLES ONTOLOGICAL CATEGORIES

A second part of the argument of the static ontology theory is that there is a "correct" ontology used by professional physicists and that for a number of key concepts that category is of emergent processes.  To this point we have given evidence that everyday thinking often and productively straddles ontological categories.

In this section we present evidence and arguments that expert reasoning straddles categories as well, and that this straddling is not "backsliding" into previous novice mindsets (as is suggested in Slotta & Chi, 2006) but is both productive and essential[6].  Concepts in physics are evolved through much effort by many researchers over long periods of time.  They do not necessarily match well with first-level embodied categories (stuff we experience directly, such as matter, processes, and states of mind).  We suggest that ontology-blending is an essential skill of an expert, necessary to deal with concepts such as energy, flow, or quantum particles that have ontological characteristics that shift with our focus and with context.

We begin with examples from the professional physics literature of how experts apply matter-based predicates to support their reasoning about non-material concepts, including processes.  We also present excerpts from a seminar delivered at the physics department of University of Maryland by a physics professor to illustrate that such conceptualization of non-material concepts using matter-based ideas is part of how expert physicists speak and reason about these concepts in professional contexts.  We then turn to a different sense in which experts cross categories, in the often explicitly contextual dependence of theoretical choices:  In different situations, experts choose matter or process based models to reason about the same physical phenomena.

## Experts Commonly Apply Matter-Based Predicates To Non-Material Concepts

We begin with two examples from the literature that discuss ontological aspects of expert reasoning in physics.  The first is in diSessa's response to Chi's and Slotta's critique (1993) of his work, where he argued that "experts, even in the most exotic corners of the field, reuse with a vengeance many reasoning patterns learned for matter" (diSessa, 1993b, p. 275):

> Conservation of field quantities is arguably the most central construct of [physics]. Everyone has heard of momentum and energy: most people have heard of 'charm' and other such quantities in accounts of modern particle physics.  These are all matter equivalents in the sense that they are conserved things whose distribution and flow in time and space are critical parameters of describing interactions. (diSessa, 1993b, p. 275).[7]

Brookes and Etkina (2006) recently noted experts' use of ontological metaphors such as *quantum well, tunneling*, and *energy levels* to reason productively about ideas in quantum mechanics.  Conceptualizing potential in a region as a *well* in which electrons are trapped, or as a *barrier* through which electrons can *tunnel* to the other side, involves a matter-like conceptualizing of the non-material concept of potential.

There are also examples from professional physics literature.  The general concept of *density*, of the intensity of some quantity at some location in space (the *amount of the quantity* per volume -- a matter-like attribute) is routinely applied to non-material entities such as energy, mass, particle number, electrical, and other currents.  Reasoning about heat flow, one of the standard examples of process used to illustrate the static ontologies view, physicists apply precisely the same equation as used to understand the diffusion of particles in a fluid.[8] Mathematically, the equation does not distinguish whether the flow is of particles or of energy (heat).  In a more generalized form, the same equation is used to understand the diffusion of information.  That is, for physicists, heat, fluids, and information can all be members of the same category of 'things that diffuse.'

There are many other examples.  In a classical treatment of electromagnetic radiation, light is well understood as electromagnetic waves.  A wave is a process in by which energy travels from one spatial location to another.  The most basic idealized form is a plane electromagnetic wave traveling through empty space and oscillating at a single frequency.  That is an idealization, because it formally represents a wave that has infinite spatial extent, but it provides the basis for models of more realistic phenomena that are limited in space and time.  A pulse or wave packet is the superposition (addition) of multiple electromagnetic waves oscillating at a small range of frequencies (Jackson, 1999, pp. 322-324).  In this way, physicists model light as a process.[9]

However, physicists routinely discuss light pulses using language that predicate analysis clearly associates with a matter-like ontology.  For example, Chu and Wong (1982) write about the *propagation* of the pulse—an example of the "movement process" predicate (Chi, Slotta, & deLeeuw, 1994, p. 40)—through a gaseous medium.  Their results discuss how the pulse becomes "distorted", and how "the main part of the pulse arrives at a signal velocity slower than c [the speed of light]".  *Distortion* and *arrive* in the strictest sense describe an object and *arrives* is an example of the "moves" attribute belonging to the matter category (Slotta, Chi & Joram, 1995; Slotta & Chi, 2006).  Consider the following passages from the same reference:

> A negative peak velocity occurs when the peak of the pulse emerges from the sample at an instant before the *peak* of the pulse enters the sample. ….
> As discussed by Garrett and McCumber and Crisp, the effect is due to a pulse reshaping, and the leading edge of the pulse is less attenuated than the trailing edge.  *The insidious aspect of the pulse shaping is that apart from an overall attenuation, the shape and width* *of the pulse can remain intact after it emerges from the sample.* [italics as in source] (Chu & Wong, 1982, p. 738).

By the methods of predicate analysis, predicates such as *emerges*, and *enters* – indicators of the attribute "moves" of ontological category of matter (Slotta, et. al., 1995; Slotta & Chi, 2006) – and references to the *shape*, *width*, *edge*, and *peak* of the pulse would indicate an object-like visualization of the pulse. Predicate analysis (Chi, 1994, Slotta and Chi, 2006) of this article reveals the use of both process and matter predicates to describe the behavior of a light pulse. We chose this article because it is well established in the literature, cited often, and the language is relatively accessible to an audience of non-physicists.

In a more recent article, article, Kim et al. (2005) talk about laser pulses interacting and coupling with matter:

> The UV probe pulse was *sent through* an optical delay line and *copropagated* with the pump pulse through the cluster jet. The probe pulse transversely *overfilled* the pump region in its focal volume. The *pump was removed* from the beam by a 800 nm dielectric mirror after interaction with the jet, and *the probe was split* and imaged onto a CCD camera to measure absorption … (Kim, et. al., 2005, p. 011201-3) [italics added]

The use of predicates such as *overfilled*, *removed*, *split*, and *sent through* would indicate a matter-like visualization of the pump and probe pulses. Note that the authors simultaneously use the predicate *copropagated*, which implies that the pump and probe pulses propagate through the same spatial region at the same time, something not possible for material objects. Also, *propagate* is a process-predicate. In this example, Kim et al. simultaneously use matter and process predicates to describe the pump and probe laser pulses.

The examples quoted illustrate that matter-like depiction of an electromagnetic pulse is used productively by experts to communicate their ideas to peers *as part of* their discussing a process-based model of a pulse. That is, just as Lakoff and Johnson describe everyday speakers as drawing on metaphors from matter-based predicates for understanding processes and mental states, so do physicists draw on matter-based metaphors to support understanding of non-material concepts.

The extension of matter-like descriptions to non-material concepts is not just restricted to the deliberate written discourse in professional scientific journals. It is evident as well in physicists' real-time interactions. To illustrate, we present an excerpt from a colloquium at the Physics Department, University of Maryland, delivered by a physicist to an audience of physicists. The speaker explains how energy is released during "magnetic reconnection" (Biskamp, 1996)[10]. His explanation relies on treating magnetic field lines as entities with spring-like tension that can form particular spatial configurations or states and energy is released as one state or spatial configuration of field lines transforms to another -- just like a compressed spring releases energy when let go:

> I will go through several slides trying to explain to you how magnetic energy gets released and I am going to use that information to show you what magnetic reconnection is all about. The first thing you need to understand is what happens when you take a bubble of magnetic field, and squash it, and then let it go -- take a bubble, squash the bubble, let it go. Well, magnetic field lines have tension just like a spring has. So, this thing wants to become round.  It's got tension in it.   At the same time, it's nearly incompressible for reasons we probably don't want to go into right now.  So, the area is preserved going from here to here (points on the slide).  Also, when you move magnetic fields around plasma sort of sticks on those field lines and moves with it.

The verbal phrases used by the speaker in association with magnetic field and field lines, such as *bubble of*, *squash it*, *let it go*, *have tension*, *become round*, *is nearly incompressible*, and *move*, are all used in conjunction with spatial manipulations of a material object.  The verbal expression in this excerpt indicates a matter-like visualization of the magnetic field, a non-material entity.  Such cross-categorical reasoning is productive for a better understanding of the phenomenon, and indeed characterizes expert reasoning rather than being an exception.

Our claim about the prevalence of cross-categorical connections in expert speech and reasoning is borne out not just in the examples that we present here but also in the very interviews that Chi and her colleagues did with physics experts.  In interviews with physics experts, Slotta and Chi (1995) focused on their reasoning about heat, light, and electric current.  Excerpts from expert interviews in Slotta's and Chi's own study (1995) show experts extending matter attributes ("moves", "supply", "quantify", and "absorb") not only to electrons and molecules but also to non-material concepts such as energy and vibrations:

"Moves":

– Expert 3: ... metal has, ah, more of a crystal lattice structure so, we can have, ah, *phonons which just travel along the lattice, moving up the lattice quicker*. Um, *phonons are really just lattice vibrations*.  [italics added] (p. 398)

– Expert 2: ... light is a form of energy that *travels faster* than other forms of energy ... It's a physical process by which *energy is transported* from one place to another.  [italics added] (p. 398)

"Supply":

– Expert 1: ... all of the *energy the stove's pumping out* is going into heating it up " [italics added] (p. 398)

"Absorb" and "Quantify"

– Expert 2: ... When I say "absorb the light," I mean some other atom, something, um, outside the emitter, somehow *takes that packet of energy* [italics added] (p. 399)

The verbal phrases or predicates italicized in the excerpts are indicators of the attributes of the matter ontology.  These experts, in the interviews done by Slotta and Chi, do not show any cognitive barrier to explanations that employ non-material lattice vibrations (phonons) and energy as entities that move from one spatial location to another or that can by bound into "packet of energy."  The topics of vibration and energy were not the target concepts being examined by Slotta and Chi, and thus were not accounted for in their analysis.

## The Ontological Choices Made By Professionals Are Context Dependent

In the above examples, the core point is that physicists are discussing non-material models but using matter-based predicates as part of their language.  There is another more explicit sense in which physicists traverse ontological categories.

As a first example, consider how a physics text introduces the topic of plasmas (ionized gasses):

> [The majority] of plasma phenomenon observed in real experiments can be explained by a rather crude model.  This model is that used in fluid mechanics, in which the identity of the individual particle is neglected, and only the motion of fluid elements taken into account…[In some other plasma problems] one has to fall back on the tedious process of following the individual particle trajectories. (Chen, 1977, pp. 45-46).

That is, depending on the situation, physicists might conceptualize a plasma as made up of fluids, or of discrete particles, and in some cases as a combination of both.  Similarly in nuclear theory, one might model a nucleus as either a "liquid drop" or as a set of neutrons and protons.  These are examples of traversing subcategories within the matter tree; other examples move again between matter and process.  With the advent of quantum mechanics, physicists now routinely choose between modeling light as comprised of waves (process) or of particles — photons.  The same is true for research in acoustics, where one might model sound as comprised of waves or of particles — 'phonons,' and for research on fundamental 'particles' such as electrons, which may also be modeled as waves.

More examples of multiple ontological representations of concepts can be found in almost all subfields of physics and would only belabor our point that physics concepts cannot be boxed in a single ontological category.  Rather, the expert use of these concepts is best described by multiple and mutually overlapping ontological views complementing each other to produce a richer and complex understanding of the physical phenomenon.

## A Specific Example Of Expert Use Of Multiple Ontologies: Blood Flow And Diffusion

The static ontologies account prominently uses the example of blood flow as an illustration of a direct process in contrast to diffusion, an emergent process (Chi, 2005; Slotta & Chi, 2006), and so we address it specifically.  In this section, we argue (i) that expert understanding of blood circulation in many contexts reflects an emergent-process view of the phenomenon and (ii) predicates from both direct and emergent process categories can apply to both blood circulation and diffusion.

*Blood Circulation is treated by experts as an emergent process.*  In many models of blood circulation, experts treat blood as a complex non-linear fluid with the dynamics of flow being determined by the Navier-Stokes equations (Ku, 1997; Li & Cheng, 1993; Taylor, Hughes, & Zarins, 1998).  There are also other models that treat blood as a collection of particles and fluid elements.  In these the particles move randomly due to local stochastic interactions that influence the equations of motion and the flow emerges from the collective particle activity (Dzwinel, Boryczko, & Yuen, 2003).

In a very simple emergent process description of blood circulation, the entities would be the blood cells and the cells of the heart muscles and those lining veins and arteries.  The rhythmic contraction and expansion of the heart, along with the random collisions of the blood cells with other identical blood cells and with cells of the heart muscle result in the movement of blood along the circulation path.  Each individual cell responds to the local pressure and interacts with the neighboring cells, but the flow-pattern emerges due to the net drift of the blood cells under this interaction.

*A description of blood flow crosses predicates across direct and emergent process categories.*  In the static view of ontologies, predicates from one ontological category cannot apply to a member of a different category.  Thus, direct-process predicates such as "constrained", "distinct", "dependent", "sequential", "terminating", and "corresponding" are said to be applicable to blood circulation; the contrasting emergent-process predicates such as "unconstrained (random)", "uniform", "independent", "simultaneous", "non-terminating", and "non-corresponding" are attributed to diffusion (Chi, 2005, p. 175).  However, for diffusion one could consider the collision of two molecules as *dependent* on the history of their collisions with other molecules; and *sequential,* in that it's much more likely that molecules at the center of a drop of dye in water will collide with other dye molecules before colliding with the water molecules.  Similarly, attributes from emergent-process categories can often be applicable to circulation:

*Simultaneous*: actions of different components such as the heart, valves in veins, movement of blood cells, movement of muscles lining arteries go on *simultaneously* during circulation.

*Unconstrained*: Almost all the different sub-components of the cardiovascular system are coupled via indirect, non-linear feedbacks (Taylor, Hughes, & Zarins, 1998).  The behavior of components is thus unconstrained in that any component can influence the activity of any other component; they are not constrained to interactions with only the components that they are in physical contact with.  For example, local changes in end organs can lead to changes in the behavior of arteries and the heart.

*Non-terminating*: The components driving the phenomenon of circulation need not terminate when the pattern of flow stops.  It is common knowledge that a heart can continue to beat for a few seconds even after being cut out of a body.  Recently, researchers have been able to grow a heart in the laboratory that exhibits the normal heart contractions without being attached to the rest of the circulatory system (Ott, H. C., et. al., 2008).

In sum, neither blood circulation nor diffusion has an inherent 'correct' ontology.  Like plasmas, electrons, waves, and vibrations, expert understanding of circulation or diffusion cannot be captured by a single ontological description or described by a single category of ontological predicates.

## Ontological Blending Is A Fundamental Component Of Modern Scientific Thought

Summarizing our discussion so far, expert knowledge and reasoning in physics crosses ontological categories, including crossing between matter and processes as well as crossing between direct and emergent processes.  It crosses these categories both tacitly and explicitly, through metaphorical connections such as in understanding conserved quantities and through explicit choices of models.

In addition to the empirical evidence cited above (and in diSessa, 1993b), there are good theoretical reasons for expecting this result.  Our sense organs and our brain's tools for interpreting the data from those sense organs are an evolutionary "satisfice".[11] They are not optimized in general, but "good enough."  For example, the rod and cone cells in our retina are only good enough to detect a very limited range of the electromagnetic spectrum.  But since the light from our sun delivers most of its radiation in that range, and since the objects we need to find to sustain us and the objects we need to avoid long enough to pass on our genes to a new generation all reflect well (are not invisible) in those frequencies, those detectors suffice (Llinas, 2003).  This does not mean that there is no information in other frequency ranges, and indeed, some animals such as bees and birds are sensitive to light we cannot see and use the information contained in them.

In a similar vein, the "natural categories" that our brain invents in order to deal with our survival are a satisfice.  They are "good enough" for everyday experience.  But the activity of science seeks understanding at a deeper and more analytic level.  Just as scientific instruments supplement our limited sensory receptors and show us unimagined structures to reality, our scientific theories supplement our limited "direct" ontologies and create more complex and subtle categories.  We are like the blind men and the elephant, striving to describe objects of more complexity than those with which we have direct experience.  The way we do this is by "mixing metaphors" – chaining cognitive blends in a context dependent way in order to get a richer and more complete understanding of phenomena that are both subtle and strange.

The emphasis the SO perspective places on emergent vs. direct processes mischaracterizes the goals and methods of science.  Science is about developing understanding and explanations for the purposes of both better describing phenomena and predicting behavior in new situations.  The primary theoretical tool for these developments is the search for *mechanism* (Machamer, 2000).  Emergence produces a mechanism, but only a reductionist one. Science seeks mechanism not only via reductionism but also on the phenomenological level.

Mechanism involves analyzing a phenomenon into component parts, establishing relationships among the parts and with their environment, and determining processes that allow chained reasoning (Russ, 2008).  Thus, we understand the floating of objects in water in terms of concepts such as density, mass, and surface tension.  While these concepts can certainly be analyzed in terms of an emergent level—resulting from interactions of atoms and molecules—it is not necessary to do so in order to do science.  Complex scientific reasoning and problem solving can be carried out with these phenomenological concepts directly, without any need for understanding their emergence.  All that is required is that we accept some parameters as given by measurement rather than by calculation from lower level quantities.  Such parameters exist at every level.

The examples of electric current and water flow that the SO theorists frequently cite as contrasting examples makes both our points clearly:  (1) Complex phenomena may require mixed ontologies; and (2) scientific mechanistic reasoning need not require emergence.

The "correct" (i.e., most reductionist) scientific explanation of current flow in circuits does not involve tiny localized electrons bouncing around at random[12] but has delocalized electrons shifting the population occupying particular states in the set of allowed quantum states. But at the macroscopic level such a detailed description is unnecessary.  No electrical engineer needs to think about electrons in deciding whether a circuit she is designing will function properly.  But she will have to think about currents and how they behave at a phenomenological level.  Here, both matter-like aspects (the conservation of current – Kirchoff's first law) and process-like aspects (the response of current to the pressure of the voltage from a battery – Kirchoff's second law) are relevant.

It is also interesting to note that a careful description of the ontology of mathematics, a fundamental tool in the ontology of modern physics[13], shows that complex mathematical structures are built up by starting with "objects" (e.g., numbers), defining "processes" that manipulate and transform those objects (e.g., functions), and then recategorizing the processes as a new kind of object (e.g., vectors in a function space), creating new processes that act on these objects, etc (Sfard, 1991). So the mixing of ontologies is not only built into the science via the phenomenology but also through the semiotic tools we use to describe and manipulate the objects and processes being considered.

## NOVICE UNDERSTANDING OF PHYSICS CONCEPTS STRADDLES ONTOLOGICAL CATEGORIES

In the previous sections, we presented arguments and evidence that (1) cross categorical reasoning productively pervades everyday speech and thought, and (2) expert reasoning about a physics concept spans multiple ontologies, sometimes between contexts and at other times within the same context.

The ease of straddling ontological categories is also reflected in student reasoning about physics concepts. Chi and Hausmann (2003) agree that novices can display an easy flexibility in ontological classification in some cases such as reclassifying whales as 'mammals' instead of 'fish', but they contend such flexibility does not extend to science concepts such as light and electric current. Our argument is in contrast to the dominant premise of SO perspective that novices are committed to matter-based reasoning for science concepts and need to undergo an extraordinary process of conceptual change in order to alter or expand their ontological commitments from matter-like to process-like. We would like to emphasize that we agree that matter-based reasoning can dominate naive reasoning in particular contexts and that it can lead to incorrect reasoning. However, as we argue, matter-based reasoning can also lead to correct reasoning and is ultimately essential for the development of expert understanding.

In the following subsections we present (i) examples from science education literature that argue against the idea of an inflexible naïve ontological view and (ii) excerpts of a student's reasoning about current in the classroom and in homework that show her switching her ontological stance without any evidence of resistance.

## Science Education Literature Provides Multiple Examples Of Novice Reasoning Spanning Multiple Ontologies

In documenting how students reason about different wave phenomenon such as propagation velocity and superposition, Wittmann observed that in a university physics class, "students approach the topic of wave physics using both event-like and object-like descriptions of wave pulses" (Wittmann, 2002, p. 97). He found that students did tend to make incorrect object-like attributions to wave pulses (e.g. thinking that a faster hand-flick would make a pulse move faster, or thinking of pulses as bouncing off of each other). But the same students could also correctly reason about waves using process resources, allowing waves pulses to pass through each other or adding up the amplitudes of two waves at the same location in space. Wittmann (2002) noted that some students transition from matter-like to process-like descriptions quickly, without noticing the inconsistencies, and he suggested that for at least some students the explanation is generated 'on the fly' depending on what aspect of the wave they most attend to in that moment (p. 113).

Moreover, some of Whitman's subject included object-like descriptions of waves pulses in their correct responses.  Reasoning correctly about the un-attenuated propagation of wave pulse one student said, "Over a long, taut spring, the friction or the loss of energy should not be significant; so the wave should be pretty much the exact same height, distance - everything." (Wittmann, 2002, p. 106)   Just like the expert reasoning in scientific literature (Chu & Wong, 1982), the student attributed matter-like *height*, and *distance* ideas to the wave, as part of her reasoning.  Second, in one particular case, the student switched from an object to a process description of wave pulses simply on being prompted (p. 112), much like Chi's and Hausmann's (2003) report of  students easily re-categorizing whales from "fish" to "mammals."

Clark (2006) investigates how students' reasoning about thermodynamics in everyday situations evolves over multiple interviews in a longitudinal study.  Heat and thermodynamics is a topic that the SO framework mentions as particularly difficult for novices because they think about heat as a material entity and heat transfer as a direct process of movement of heat or of hot molecules (Reiner, Slotta, Chi, and Resnick, 2000; Slotta and Chi, 2006).  Clark (2006) follows the progress of a cohort of students, interviewing them over an eighth grade semester followed by additional interviews for four selected students before the start of tenth and twelfth grades.  While Clark's interviews do show that novices often use matter-based reasoning, it also shows the persistence of multiple ideas, at times conflicting, over extended periods (p. 545).

We consider the ideas expressed by Luis during the fourth interview (p. 504), conducted in the eighth grade semester.  Many of Luis's explanations of thermodynamic phenomenon treated heat as a substance, just like those of the novices in the Slotta's and Chi's study (1995).  For example, while explaining conduction, Luis uses the idea that (p. 504) "Metal is a good conductor and so heat travels through it easily[14]."  Here, Luis employs the matter-like "move" predicate in conjunction with heat.

In that same interview, however, when explaining how ice cream warms up, Luis correctly uses the idea that (p. 504) "Atoms in ice cream move faster when they come in contact with atoms in the air which makes the ice cream warm up."  In the context of this latter explanation, Luis seems to employ the emergent-process ontology where heat transfer is the process by which slower atoms gain energy via interaction with more energetic atoms and start to move faster.  The context dependent variation in the ontology underlying Luis's explanation for heat transfer, even within a single interview, seems to counter the idea that as a novice Luis is tied to a single ontology for a concept or that conceptualization of heat as a process necessarily requires him to suppress the alternative matter-like characterization.

There are also moments in which Luis's reasoning does not seem restricted to either one of the matter or process categories.  For example, during the first interview (p. 502), Luis explains that ice makes a nail cold with the idea that "the "atoms of ice" are "sending cold to the nail and making (the nail's) atoms move slower" which makes the nail feel cold[15].  Here, while Luis talks about cold as a material entity (indicated by the use of the matter attribute "move"), he also displays an understanding of heat as related to vibration of atoms in the substance, a process like view (Chi, 2005, p. 189).  Be it within context or between contexts we are hard pressed to assign Luis's understanding of heat and heat transfer to a single ontological category or to attribute his use of matter-predicates in association with heat as resulting from a lack of the emergent process category or a cognitive barrier to associating heat with the process category.

Atkins (2004) presented a case study of a fourth-grade class reasoning about what would happen to the water in a cup, if the cup were falling upside-down:  Would the water fall out of the cup?  One student, Miranda, generated an analogy to her experience of swinging a small toy cat in a basket over her head in a vertical circle.  The cat stayed in the basket, which made her think (correctly) that the water would stay in the cup.  Atkins argued that in generating the analogy, Miranda was effectively generating an *ad hoc* category, with members including her swinging cat-in-basket and the falling water-in-cup.  This was creative, Atkins noted, in the sense Chi described, of "re-representing an entity or a situation from one 'ontological' tree of concepts and categories to another ontological tree of concepts and categories" (Chi, 1997, p. 209).  Within the static ontologies framework, the liquid water and the solid cat belong to parallel subcategories of matter; it should not make sense to put them together, except in the context of predicates that apply to all matter uniformly.  There was no evidence, however, that Miranda or any other student found the comparison difficult to understand; more at issue was the difference in the motions of the containers than the difference in the substances they contained.

In a recent paper, Levy and Wilensky (2008) explicitly address the issue of students' ability to use reasoning from emergence in the description of complex phenomenon in the context of situations familiar to the students such as the process of a group of students scattering for calisthenics and spreading of rumors among students in a class.  The authors observed that all of the ten sixth-grade students in their study argued the phenomenon from the microscopic (emergent-process like) and the macroscopic level (direct-process like) and constructed a mid-level description to bridge the reasoning from one level to the other.  To varying degrees *all* of ten students interviewed by Levy and Wilensky used complex-system resources such as decentralized control, unpredictability, distinction between the levels (microscopic interaction rules and the larger group behavior), equilibration process and small change-big effect for reasoning through the task.  Decentralized control, unpredictability, distinction between levels, and equilibration process correspond to emergent process attributes and use of these resources would serve as an indication that novices are familiar with the emergent-process category.  As an example, consider the authors' analysis of the student Jessica's reasoning about how students scatter in a room:

> Thus, Jessica divides the process by which the class scatters into two parts. In the first part, rows form emergently, an agent-based from of reasoning. In the second part, the rows spread out in a staggered process, via centralized control and local rearrangement. (Levy & Wilensky, 2008, p. 26)

Here, agent-based reasoning refers to reasoning where the interactions of entities at one level lead to the emergence of group-level properties without direct correspondence between the actions/interactions of agents and the emergent property.  Thus agent-based reasoning directly refers to reasoning in terms of emergent-processes.  In contrast, centralized control reflects direct-process ontology.  Levy and Wilensky note that Jessica's explanation involves aspects of emergent-process reasoning as well as aspects of direct-process reasoning - categories that are considered incommensurate by Chi (2005).

Levy and Wilensky show that in familiar contexts, novices not only make use of resources commonly used by experts to understand emergent phenomenon but they can also simultaneously coordinate resources in two different ontologies for the same phenomenon.  The authors explicitly argue against the distinctness of ontological categories in the SO view:

[Our results do not] provide support for the incommensurability view. While we have seen evidence for Chi's (2000, 2005) described difficulties in bridging between these two forms of reasoning [referring to direct and emergent processes], we have decidedly not found the two forms to be unbridgeable. Some of the students did connect the two forms of reasoning. Doing so through mid-level construction is important for their sense making of complex systems. (Levy & Wilensky, 2008, p. 42)

Many more examples of multiple ontologies underlying students' reasoning can easily be found in science education literature where extensive interview data is made available, irrespective of whether the original studies were intended to study ontological commitments. We have presented four such examples and more would just belabor the point. In the following sub-section, we present a case study of a student reasoning about electric current in the physics classroom.

## Spontaneous Ontological Transitions And Matter-Based Ontology Of Electric Current Is Productive In Student Reasoning

We consider the specific example of electric current, a topic that has been central to SO arguments (Slotta, Chi & Joram, 1995; Slotta & Chi 2006) and consider a case study of a particular student, whom we will refer to as Kimberly. The data comes from videotapes of an inquiry based physics course at the University of Maryland. This was a 4-credit, lab course for elementary education majors, most of who had not taken high-school physics. The instructor was a physics professor and the first author was the teaching assistant. The course was divided into three units, each roughly four weeks long. The class was driven by student ideas and experimental observations. There was no textbook, only a lab manual that had minimal instructions to guide experiments and no explanations (Layman, 1996).

In the episodes and data we consider from this class, we adopt SO theorists' predicate analysis to identify ontological categorizations. Recall the example above in which a student's applying predicates that described "current as a substance-like entity with attributes of (a) 'moving,' (b) 'can be consumed,' (c) 'can be quantified,' and (d) 'moves,'" (Slotta and Chi, 2006, p. 265) constituted evidence the student views current as a member of the matter-ontology. If, on the other hand, a student were to use process predicates, describing current as "propagating," as "movement of electrons," or as "taking place," it would constitute evidence she views current as a process.

The students were investigating electric current using batteries, bulbs, and wires. On the day of this episode, they were trying to understand how multiple batteries and multiple bulbs affect current in a series circuit. They had already made some observations of the brightness of bulbs as they added more batteries to the circuit (keeping the number of bulbs constant) or as they added more bulbs (keeping the number of batteries constant). At one point, the instructor asked if the current at one end of a particular arrangement of batteries is always the same. Here is how one student, Kimberly, answered that question:

Kimberly: Well, I am not sure if this is true but, when we added more light bulbs you saw they got dimmer, so they're kind of *sharing* what the current for one light bulb would be. So *it got like divided* between bulbs. So maybe, current isn't measured by the number of batteries. Maybe it's measured by like the number of batteries and then *split up by what's it being used by*.

Less than an hour later, the class was discussing whether, in a series circuit with multiple bulbs, the current is stronger just outside the battery and gets divided when it reaches the bulbs, or whether is it same throughout the circuit.  Kimberly's reasoning looked different:

> Kimberly: Maybe the electrons...I don't know. Don't hold me to this. But like think of them as a string of electrons; like they are all connected. So I guess everything else is connected but its going, like, through, like, a tunnel through it. 'cause, like, If it's a conductor it allows them to flow through.  If you pull that string through the whole thing, they are all going to go at the same pace - its not gonna slow down at one point and get fast, or else the string will actually break.

In both discussions, Kimberly argued that the electric current would depend on the number of bulbs in the circuit, but her reasoning had clearly different flavors to it.  In the first discussion, Kimberly used matter-like expressions, *sharing*, *got divided*, *split up* and *being used by* to depict the idea that current in a series circuit is lessened if you add more bulbs.  From the SO list of predicates (Chi, Slotta & de Leeuw, 1994, p. 40), *sharing*, *got divided*, and *split up*, are verbal indicators for "equivalent amounts" predicate; *being used by* indicates the "consume" predicate.  "Equivalent amounts" and "consume" are both matter predicates, and would signify that Kimberly, in the first utterance, categorized electric current ontologically as matter.  In this second instance, Kimberly's explanation reflected an understanding of current as related to how fast electrons are moving and that the current is the *same throughout* the circuit - indicators of "movement process" and "system-wide" process predicates (for lists of matter and process predicates see Chi, Slotta & de Leeuw (1994) and Slotta & Chi (2006)).  By the SO methodology, her use of process predicates in the second instance would indicate a process ontology of electric current.  Thus, analysis of her verbal reasoning using predicate analysis indicates that Kimberly had different ontological stance on current in these two instances within the same hour.

The students also had to hand in a weekly assignment that contributed in part towards their grade for the course.  Kimberly's response to a homework question that was assigned during this week demonstrates this same understanding of current as a process:

> One model for electricity that is evident in everyday life is the "Heart Model." … These blood cells flow through the veins and arteries much like electrons do through the wires. These are the transportation for the electrons. … The current in my model is the blood-flow or blood pressure, how fast the blood is being pushed through the channels, the veins and arteries. To measure current in this model one would have to count the number of blood cells per second that passes through one point of this path, in essence the circuit.

The blood model has been of particular concern from the SO perspective as likely to promote either matter-based or direct-process-based categorization (Slotta & Chi, 2006).  Here, Kimberly explicitly mapped current to the idea of flow in her model using the "movement process" attribute in association with current.  In other words, in this context, she views current as the movement of electrons (for electric current) and blood cells (for blood current); she does not treat current as an entity that moves. There was no evidence she was constrained to a matter-based ontology. There is evidence she was reasoning about flow as a direct-process—she is thinking about the flow as caused directly by the translation of blood cells/electrons; an emergent picture would describe the flow as resulting from a net drift even when individual particle motion was random or non-directed. But this helped her to make sense of several observations, including of current being the same at every point in a series circuit, of the effect of adding bulbs or batteries, and the role of directionality of batteries:

> Our results so far relate to this model in that we have found that current is constant throughout a circuit. In the heart model, blood cannot go faster at some points then others because there is only a limited space to flow through.
> The bulbs act as resistance to the current or flow. In the heart model, the more places the flow needs to pass through the slower the blood will flow through that path.
> The fact that multiple batteries in the same direction make current stronger and faster is proven by the fact that if someone had two hearts pumping in the same direction the blood would flow two times as fast. However if the hearts pump in opposite directions with equal strength it then is at a standstill and there will be no evident current.

She even used this model to argue for the ontology of the battery as a pump rather than a source of electrons in the circuit:

> Some people asked if electrons were always there and when hooked up to a battery if the battery provided the electrons or just acted as a pump. The heart model suggests that the blood would still be floating throughout the body but would not have any motion, and therefore says that electrons are everywhere and that the battery just helps to push them along.

It might seem that Kimberly had a substance-view of current at the beginning of the class and that by the end of that class she had formed a process-based understanding of electric current, as indicated by her responses later in the hour and from the homework. However, while she does talk about current as a flow, she also talked about "multiple batteries making the current faster" in the homework, treating current as the thing that moves. Also, analysis of more of her homework does not indicate a static ontological stance. Her previous answers on the same homework contain segments in which she refers to current 'getting divided' by the number of bulbs and current 'flowing through' the circuit. More interesting than those little bits of language is her reasoning about parallel circuits on the *following* week's homework:

> In order for two bulbs in the same circuit to have different amounts of current flowing through them…Current we have defined as the number of electrons passing through the bulb per second…The more electrons that pass per second, the brighter the bulb and fewer electrons that pass the dimmer the light of the bulb…however, we had a very "heated" discussion on how to determine the amount of current that flows out of the batteries before it breaks out to the two different paths available to it. We believe to have determined that the batteries produce the current to two "separate series circuits" in this case. Since path 1 had four batteries worth of current, and path 2 has two batteries worth of current, we then determined the current coming out of the batteries is therefore six batteries worth. [emphasis as in source]

Here, Kimberly seems to retain the idea of current as the movement of electrons but she also uses matter-like predicates such as 'flows out of the battery' (corresponding matter-attribute—"moves"), 'it breaks out into two paths' (corresponding matter-attribute—"moves"), and 'batteries produce the current' (corresponding matter-attribute—"supply"). This reasoning supports her understanding of the amount of current as conserved while dividing into two paths.

In Section I, we argued and gave evidence that everyday reasoning easily crosses ontological categories. Kimberly's case illustrates a student's reasoning crossing categories specifically with respect to electricity. Given the extent to which everyday reasoning straddles ontological categories, it would be strange for it not to happen in novice reasoning about novel phenomena.

More important, Kimberly's reasoning based on the "wrong" ontological categories was productive for her learning about electricity. Her reasoning about current as a direct process, by analogy to blood flow, supported her explanations of series circuits with respect to measurements of current at different points, the effects of varying the number of bulbs or batteries, and the effects of different orientations of batteries, all in ways that fit the evidence and that physicists would recognize as correct. Her reasoning about current using matter predicates supported her explanation of what happens in a parallel circuit, specifically with respect to the conservation of current as it splits between two paths —reasoning physicists recognize as Kirchoff's Current Law. From Lakoff's and Johnson's (2003) account, we expect it is generally the case that understanding any quantity as conserved has metaphorical roots in our first matter-based experiences of conservation.

Thus Kimberly's reasoning drew from different ontological categories, with no apparent difficulty, and it did so productively. An instructional strategy aimed at divorcing her from these reasoning resources might hinder her understanding of the phenomenon rather than promote it.

## CONCLUSIONS AND IMPLICATIONS

In the preceding three sections, we reviewed evidence that ontological flexibility is pervasive in everyday discourse, provided case-study existence proofs of that flexibility in experts' and novices' reasoning in physics, and gave arguments to suggest that the case studies are not unusual. In our experience as physicists and as physics instructors, these examples are typical, although we have not conducted any formal studies to establish that typicality.

From our perspective, the evidence overwhelmingly belies the SO view that ontological categories are distinct, stable, and constraining. Rather, we contend, an individual's ontological reasoning can respond dynamically to the immediate needs of the moment. As we illustrated, an expert may reason about a process as though it is *matter*, when it is effective and efficient to do so, then move back to treating it as a *process* as needed. A novice may move back and forth as well. Moreover, we illustrated, an individual's reasoning in any particular moment may not fall neatly into a single ontological category. Mixed-ontologies are a fundamental part of the thinking of professional scientists, and mixing ontologies is common and productive in both everyday speech and in the behavior of naïve students.

## Implications For Research Methodology: Explaining Data From Controlled Experiments From A Dynamic Perspective

In our view, previous research has identified fixed categories precisely because the methods involved carefully controlled contexts. Within those contexts, students' reasoning fell into stable patterns that researchers observed. We do not doubt that in particular circumstances, ontological reasoning can be stably constrained to a single category, and we agree with SO theorists that in such moments the wrong stability can lead students to incorrect conclusions. But we do not interpret such stability as a fixed property of cognition.

Dunbar and Blanchette (2001) made this argument powerfully in their research on analogical reasoning, showing how "*in vitro*" studies of students' abilities for analogy produced quite different results from their *"in vivo"* studies in naturalistic settings. Here, the difference in theoretical perspectives between a dynamic or static account of ontological knowledge has similar implications for methodology. Naturalistic case studies, such as those in the examples above, can reveal more complex dynamics than controlled experimental conditions, especially those designed based on a static ontologies framework. We have begun to explore controlled experimental conditions designed specifically to cue multiple stabilities in ontological reasoning. For example, we have seen preliminary evidence that graduate students in physics will reliably draw on a matter-based metaphor for heat, when asked to explain how coffee cools in a ceramic mug, but that they will as reliably shift to a process-based metaphor when asked to explain the mechanism of heat transfer (Gupta, Hammer, & Redish, in preparation).

## Dynamic Perspective Emphasizes The Development Of Skills To Evaluate Productiveness Of Multiple Descriptions

Of more immediate importance, as we have noted, the difference between a static and a dynamic view of naïve ontologies has strong implications for instruction. The SO view implies that instruction should inhibit reasoning based on the "wrong" ontological category. We have argued that this would hinder students' learning and, if it were possible to impose at professional meetings, expert reasoning. To restrict thinking and language about optical pulses to predicates in a process category would introduce barriers to thinking about reshaping the pulse, pulse distortion, and pulse absorption that the articles we quoted discuss. Moreover, an instructional strategy focused on ontological change could undermine students' applying the resources they bring to the table. For example, forcibly divorcing the concept of current from the matter category could prevent a student like Kimberly from understanding Kirchoff's Current Law. As well, instructional policing of students' thought and language for "correct" ontology could exacerbate existing difficulties of students experiencing the discipline of physics as disconnected from their everyday knowledge of the world (Hammer, 1994; Redish, Steinberg & Saul, 1998).

The dynamic reasoning exhibited by experts in the examples above indicates that being aware of how concepts can be treated in multiple ways depending on the context is an important part of expert epistemology in physics. The emphasis on a single correct description of a concept and, more importantly, on avoiding all alternate descriptions of it during instruction seems to undermine this expert epistemological stance. Rather, instruction should encourage students to explore multiple descriptions of a phenomenon and help them evaluate the appropriateness and productivity of each description in different contexts.

Students bring with them a variety of resources to draw on for reasoning about matter, as well as about events and processes. Instruction should be geared towards tapping these resources, drawing students' attention to the different descriptions of natural phenomena, and discussing when some description is appropriate and when it is not. Students' developing expertise means their becoming aware of the productiveness and limitations of the descriptions they use and the resources that they have at hand.

## SUMMARY

One of the principal theoretical questions facing researchers is the nature and organization of knowledge as cognitive structure. Many researchers have focused on conceptual knowledge in characterizing cognitive structure (Carey, 1986; diSessa, 1993a, Strike & Posner, 1992). The ontological view can add significantly to this issue since it provides information on the intuitive categorization of conceptual knowledge. The prominent view in this direction has been that of a static model of ontological knowledge; that the intuitive categorization of science concepts by experts as well as novices is fixed across contexts and requires a conceptual change for alteration. In this paper, we argue against this static ontology for physics concepts. Our analysis of verbal predicates following Slotta's and Chi's (2006) methods shows both experts and novices moving across ontological categories for the same concept. This evidence points towards a dynamic picture where our ontological knowledge is flexible and ideas in the world and ontological categories are multiply connected. Theoretically, this suggests that our conceptual knowledge organization is likely to be network-like rather than hierarchical.

How can we then describe the ontological knowledge of a person? The greatest number of entities we see around us are *matter*-like; so it is little surprise that even children have a well-developed *matter* category. But adults as well as children display a wider array of resources and beliefs about the ontology of the world. We have productive ways to think about the ontology of *events* such as a footrace and *processes* such as the spreading of news. In many cases we think of the ontology of an entity as real (for example, it is reasonable to assume that the chair I am sitting on really is a solid object) while in other cases we understand the ontology of an entity as imagined/un-real (for example, we understand Tom Sawyer to be a boy; at the same time we do not consider him to have been a "real" boy but consider him a book character). In numerous examples, especially for imagined/unreal entities, we easily allow them to have an ambiguous ontology or a patchwork of attributes from more than one category forming a unique ontological category. Thus, a centaur is taken to be a man and a horse at the same time and ents in *The Lord of the Rings* (Tolkien, 2004) were sentient trees that could walk and talk. All of these, one's knowledge of more and less prominent categories, and beliefs and attitudes towards ontology, can be thought of as cognitive resources that affect the ontology underlying our thoughts and reasoning. In future studies, we aim to characterize the form of ontological knowledge and development of new ontological resources.

Instructionally, the dynamic view suggests building from students' everyday resources. We need to devise instructional strategies to harness, in science-learning contexts, resources that students activate in other contexts; and this is not restricted to thinking in terms of *matter* and *processes* but also tapping into their beliefs and attitudes towards ontology.

In our view the difference between experts and novices reasoning in these respects is not entirely explained by differences in their ontological resources. Rather, it results in part from differences in their meta-cognitive awareness and control of those resources. Different epistemologies could elicit very different reasoning resources that in turn could modify an ontological view.  Future research programs should aim at modeling the dynamic interaction of epistemological and conceptual resources.

In summary, we have argued that expert as well as novice reasoning about entities and concepts in science is context-dependent and both groups borrow from multiple ontological categories in a dynamic fashion.  This challenges the static model of naïve ontology.  In effect, we are making the case for an ontological shift in the way that researchers construe and model cognition and reasoning.  Instead of modeling the patterns in students' reasoning as *directly* corresponding to, or materially constrained by, equivalent structures in the mind, we are arguing for developing *emergent-process* models of the phenomenon—where the dynamics of interconnected knowledge elements can give rise to contextual variability as well as robust patterns.  Dynamic models can explain a larger set of observations of cognitive behavior.  The development of such models would require theoretical as well as methodological changes to the current approach and present theoretical and methodological challenges to be solved.

## ACKNOWLEDGEMENTS

We would like to thank the members of the Physics Education Research Group at the University of Maryland, College Park for useful discussions.  We would also like to thank James Drake, Department of Physics, University of Maryland, College Park, for permission to include an excerpt from the seminar he gave as part of the departmental Colloquium Series and Carol Alley for providing the video recording of the event.  We are thankful for helpful comments from Iris Tabak as well as three anonymous reviewers.  This research was supported in part by funds from NSF grants REC 04-40113 & DUE 05-24987.  The views expressed in the article are the authors' own and do not necessarily reflect those of the National Science Foundation.

[1] It is an expositional difficulty that there is no clear term by which to refer to the superset of all possible objects of thought, objects in the grammatical sense — the word *objects*, for example, clearly suggests the category of *matter*. Here we follow Chi (1992, p. 131) and use the term "entities," although that term as well connotes matter specifically.

[2] To the SO theorist, the mutual exclusivity of categories means that concepts belong to one and only one category at any level of the hierarchy. This approach is reminiscent of classical categorization theory. As we discuss in section III, modern categorization theory, both in psychology (Sternberg & Ben Zeev, 2006, Ch. 3) and in linguistics (Evans & Green, 2006, Ch. 8) is much more fluid and less constrained.

[3] 'Constraint-based interactions' (Chi and Slotta, 1993) was later renamed as 'emergent processes' (Chi, 2005).

[4] In this respect, Chi and Hausmann (2003) differs from the SO perspective. They acknowledge that metaphors in everyday speech reflect flexibility of ontological reasoning. Still, however, they maintain that "attributes of [different ontological] trees are totally, mutually exclusive" (Chi and Hausmann, 2003, p.6), and the possibility of ontological flexibility is not discussed elsewhere in the SO literature we cited above.

[5] See references in Sternberg & Ben-Zeev, 2001

[6] In general, through the development of the ontology framework, Chi and her colleagues have given greater weight to discussion of novices' ideas than to those of experts. Where they have discussed experts in more detail (Slotta, Chi, and Joram, 1995; Chi and Hausmann, 2003), the basic argument has been that experts think about science concepts such as light, heat, and electric current as processes in general and emergent-processes in particular. They discussed

ontological shifts in expert reasoning in cases where (1) an expert in everyday context could use matter-based attributes for a science concept and (2) examples from the history of science show that certain breakthroughs in science can be understood as ontological category shifts. However, we engage in a detailed discussion of how experts think about science concepts in the current scientific professional literature and professional seminars, because it is important to characterize what expertise looks like when experts are reasoning about science concepts with other experts.

[7] Conservation is not explicitly listed as a matter-like attribute within the SO literature, but it is a resource that derives from our basic experiences with objects in the real world. A toy or block left at a place remains there. When moved from place-A to place-B, it is the same toy/block that is no longer present at place-A but is now available at place-B. These repeated experiences in the world in which a child is immersed form the basis for the development of ideas about object permanence, motion, and spatial location. These can then be thought of as the seeds for the development of the idea of conservation: that objects (or matter, in general) does not just disappear or appear out of nowhere. In this sense, conservation, is a matter-like resource, closely aligned to *spatial location* and *moves* matter-like predicates of the SO model.

[8] The diffusion equation, $\frac{\partial \varphi}{\partial t} = \kappa \nabla^2 \varphi$ , is used to describe the movement of particles as well as of heat. $\varphi$ can represent either temperature or the concentration of a substance at a given point in space. The equation essentially expresses the idea that heat flows more quickly from one place to another if the change in temperature is greater, and the same is true of particles. Our point is that *the same reasoning applies* to immaterial heat as to material particles.